\documentclass[onecolumn,preprintnumbers,preprint,superscriptaddress,11pt,floatfix,nofootinbib]{revtex4}
\pdfoutput=1

\usepackage{graphicx}%
\usepackage{dcolumn}
\usepackage{bm}

\usepackage{amsmath}
\usepackage{amsfonts}
\usepackage{slashed}

\newcommand{\eqnref}[1]{Eq.~(\ref{eqn:#1})}

\newcommand{\secref}[1]{Sec.~\ref{sec:#1}}

\newcommand{\appref}[1]{Appendix~\ref{sec:#1}}
\newcommand{\figref}[1]{Fig.~\ref{fig:#1}}

\newcommand{\tableref}[1]{Table~\ref{table:#1}}

\begin{document}

\preprint{UCI-TR-2012-16, FERMILAB-PUB-12-560-PPD, LA-UR-12-25423}

\title{Compatibility of $\theta_{13}$ and the Type I Seesaw Model with
  $A_4$ Symmetry}

\title{Compatibility of $\theta_{13}$ and the Type I Seesaw Model with
  $A_4$ Symmetry}

\author{Mu-Chun Chen}
\affiliation{Department of Physics and Astronomy, University of
California, Irvine, CA 92697, USA}
\author{Jinrui Huang}
\affiliation{Department of Physics and Astronomy, University of
California, Irvine, CA 92697, USA}
\affiliation{Theoretical Division, T-2, MS B285, Los Alamos 
National Laboratory, Los Alamos, NM 87545, USA}
\author{Jon-Michael O'Bryan}
\affiliation{Department of Physics and Astronomy, University of
California, Irvine, CA 92697, USA}
\author{Alexander M. Wijangco}
\affiliation{Department of Physics and Astronomy, University of
California, Irvine, CA 92697, USA}
\author{Felix Yu}
\affiliation{Theoretical Physics Department, Fermi National
Accelerator Laboratory, P.O.~Box 500, Batavia, IL 60510, USA}
\normalsize
\date{\today\\[1.75cm]}

\begin{abstract}
We derive formulae for neutrino masses and mixing angles in a type I
seesaw framework with an underlying $A_4$ flavor symmetry.  In
particular, the Majorana neutrino mass matrix includes contributions
from an $A_4$ triplet, $1$, $1'$, and $1''$ flavon fields.  Using
these formulae, we constrain the general $A_4$ parameter space using
the updated global fits on neutrino mixing angles and mass squared
differences, including results from the Daya Bay and RENO experiments,
and we find predictive relations among the mixing parameters for
certain choices of the triplet vacuum expectation value.  In the
normal hierarchy case, sizable deviation from maximal atmospheric
mixing is predicted, and such deviation is strongly correlated with
the value of $\theta_{13}$ in the range of $\sim (8-10)^{\circ}$. On
the other hand, such deviation is negligible and insensitive to
$\theta_{13}$ in the inverted mass hierarchy case.  We also show
expectations for the Dirac $CP$ phase resulting from the parameter
scan.  Future refined measurements of neutrino mixing angles will test
these predicted correlations and potentially show evidence for
particular triplet vev patterns.
\end{abstract}

\maketitle


\section{Introduction}
\label{sec:intro}

With the independent discoveries of a nonzero $\theta_{13}$ from the
Daya Bay~\cite{An:2012eh} and RENO~\cite{Ahn:2012nd} collaborations,
and the supporting hints from the T2K~\cite{Abe:2011sj},
MINOS~\cite{Adamson:2011qu}, and Double Chooz~\cite{Abe:2011fz}
experiments, we now possess the first complete experimental picture of
the Pontecorvo-Maki-Nakagawa-Sakata (PMNS) mixing matrix.  Following a
recent global analysis of neutrino oscillation parameters from
Ref.~\cite{Tortola:2012te} (see also~\cite{Fogli:2012ua,
  Morisi:2012fg,GonzalezGarcia:2012sz} and 
\cite{Machado:2011ar, Schwetz:2011zk, Schwetz:2011qt}), we can summarize the experimental status
to date as
\begin{eqnarray}
\label{eqn:exp_results}
\sin^2 \theta_{12} &=& 0.320^{+0.015}_{-0.017} \ , \\
\sin^2 \theta_{23} &=& \left\{
\begin{array}{c}
0.49^{+0.08}_{-0.05} \quad \text{(Normal)} \\
0.53^{+0.05}_{-0.07} \quad \text{(Inverted)} \\
\end{array} \right. \ , \\
\sin^2 \theta_{13} &=& \left\{ 
\begin{array}{c}
0.026^{+0.003}_{-0.004} \quad \text{(Normal)} \\
0.027^{+0.003}_{-0.004} \quad \text{(Inverted)} \\
\end{array} \right. \ , \\
\Delta m_{21}^2 &=& 7.62 \pm 0.19 \times 10^{-5} \text{ eV}^2 \ , \\
\Delta m_{31}^2 &=& \left\{
\begin{array}{c}
2.53^{+0.08}_{-0.10} \times 10^{-3} \text{ eV}^2 \quad \text{(Normal)} \\
-(2.40^{+0.10}_{-0.07}) \times 10^{-3} \text{ eV}^2 \quad \text{(Inverted)} \\
\end{array} \right. \ . 
\end{eqnarray}

Before the $\theta_{13} \neq 0$ discovery, the PMNS matrix was
consistent with the tribimaximal (TBM) mixing pattern~\cite{Harrison:2002er}, 
which can be written as
\begin{equation}
\label{eqn:TBM}
U_{TBM} = \left( \begin{array}{ccc}
\vspace{4pt}
 \dfrac{2}{\sqrt{6}} &  \dfrac{1}{\sqrt{3}} & 0 \\
\vspace{4pt}
-\dfrac{1}{\sqrt{6}} &  \dfrac{1}{\sqrt{3}} & \dfrac{1}{\sqrt{2}} \\
\vspace{4pt}
 \dfrac{1}{\sqrt{6}} & -\dfrac{1}{\sqrt{3}} & \dfrac{1}{\sqrt{2}} \\
\end{array} \right),
\end{equation}
where we have adopted the same phase convention as the
PDG~\cite{Nakamura:2010zzi} for placement of the minus signs.  The TBM
mixing pattern gives the solar mixing angle corresponding to $\sin^2
\theta_{12} = 1/3$, the atmospheric mixing angle $\sin^2 \theta_{23} =
1/2$, and reactor mixing angle $\sin^2 \theta_{13} = 0$.  Neutrino
mass matrices that give rise to TBM mixing have distinct invariants
that can be traced to discrete symmetries such as the Klein $Z_2
\times Z_2$ group or the symmetry group $S_4$~\cite{Lam:2008rs,
  Lam:2008sh, Grimus:2009pg, Lam:2009hn} (see also,
\cite{Hernandez:2012ra}).  On the other hand, by introducing dynamical
(flavon) fields, the TBM mixing pattern can arise from a smaller
underlying finite group, the tetrahedral group, $A_4$.  To be
compatible with grand unification, a successful generation of
appropriate lepton and quark masses and mixing angles can be achieved
by considering $T^{\prime}$, the double covering group of $A_{4}$. In
an $SU(5)$ grand unified model~\cite{Chen:2007afa}, the $T^\prime$
group also affords a novel origin of $CP$ violation from complex
Clebsch-Gordan coefficients~\cite{Chen:2009gf}, while in a
Randall-Sundrum model~\cite{Chen:2009gy, Chen:2009hr}, the $T^\prime$
flavor symmetry is simultaneously used to forbid tree-level
flavor-changing neutral currents (FCNCs).  For reviews of the status
of $A_4$ and $S_4$ models, the tribimaximal and bimaximal paradigm,
see~\cite{Altarelli:2010gt, Altarelli:2012bn, Altarelli:2012ss}. For early work in mixing and GUT theories, see~\cite{Ramond:2003kk}.

The literature on the interrelation between TBM neutrino mixing
matrices and finite group flavor symmetries is vast.  With discovery
of a nonzero $\theta_{13}$ from Daya Bay and RENO, however, the TBM
prediction for a vanishing reactor angle is ruled out, calling into
question the entire TBM paradigm.  We emphasize, however, that the
underlying flavor symmetries that naturally give rise to TBM mixing
are nevertheless viable options for explaining the updated PMNS mixing
pattern.  In particular, we demonstrate that $A_4$, when we include
flavons that are not included in the usual TBM analysis, can readily
accommodate a large value of $\theta_{13}$ and retain predictivity for
$\delta$, the PMNS $CP$-violating Dirac phase.

Work before the Daya Bay and RENO results that focused on finite group
symmetries and generating nonzero $\theta_{13}$
includes~\cite{Altarelli:2005yp}, which studied corrections to TBM
from higher dimensional operators and~\cite{Antusch:2005gp}, which
discussed renormalization group (RG) equations in see-saw models.
More recent work looking to use higher dimensional operators to
generate violations of the TBM scheme or neutrino phenomenology
include~\cite{Ahn:2011pq, Ahn:2012tv}.  The authors
of~\cite{Cooper:2011rh} conclude higher dimension operators and RG
effects are equally important for leptogenesis.
In~\cite{Boudjemaa:2008jf}, it was found that the size of corrections
to neutrino mixing sum rules coming from renormalization group running
are small.  Correspondingly, in~\cite{Luo:2012ce}, it was shown that a
large $\theta_{13}$ value cannot be generated through running if
$\theta_{13}$ starts at 0.  In~\cite{Antusch:2010tf}, NLO and NNLO
expressions were given for neutrino mixing angles in hierarchal mass
scenarios with sequential dominance.

Nevertheless, we note that it is possible, with the so-called Hilbert
basis method~\cite{Kappl:2011vi}, to construct supersymmetric models
where these higher dimensional operators in the holomorphic
superpotential vanish.  On the other hand, certain flavon-induced
corrections in the non-holomorphic K\"ahler
potential~\cite{Leurer:1993gy} cannot be forbidden by {\it any}
conventional symmetries.  Hence, even in the original $A_4$ models
where $\theta_{13}$ vanishes at leading order, once these K\"ahler
corrections are properly taken into account, a sizable $\theta_{13}$
can be attained that is compatible with the current experimental
value~\cite{Chen:2012ha}.

Much of the recent literature has focused solely on accommodating
non-zero $\theta_{13}$ with low-energy perturbations to the TBM matrix
and ignore or only briefly discuss the possible underlying UV physics
responsible for these deviations.  In this vein, one popular
parametrization was introduced in~\cite{King:2007pr}, which introduced
deviations from TBM values using $s$, $a$, and $r$ for the solar,
atmospheric, and reactor mixing angles, respectively.  Further work
along this line, such as parameter scans in this space of deviations,
include~\cite{King:2009qt, Ahn:2011if, Araki:2011wn, Chao:2011sp,
  Zheng:2011uz, Grinstein:2012yu, He:2011gb, Cooper:2012wf, Ma:2012zm}
and also~\cite{Eby:2011aa, Chu:2011jg, Benaoum:2012my, Roy:2012qm},
which used a similar approach but an equivalent set of parameters.
One drawback of such approaches is that complete UV flavor models
generally predict additional flavor violating effects, which are not
captured in these low-energy deviation
studies~\cite{Varzielas:2010mp}.  Separately, TBM and BM mixing
scenarios were studied for the case of nearly degenerate neutrino
masses where loop corrections provide large effects in mixing
angles~\cite{Araki:2011qy}.  Loop corrections to the Type I seesaw
Majorana neutrino mass matrix leading to nonzero $\theta_{13}$ were
also studied in~\cite{Brahmachari:2012cq}.  Studies of generating
nonzero $\theta_{13}$ from breaking $\mu$-$\tau$ symmetry
include~\cite{Liao:2012xm, Zhou:2012zj, Xing:2012sj}.

In this work, we adopt a top-down approach where we will constrain the
UV parameter space by the low energy neutrino observables. While other
$A_4$ studies may have looked at the additional effects from $1'$ and
$1''$ flavons contributing to the Majorana neutrino mass matrix,
including~\cite{King:2011zj, King:2011ab, Shimizu:2011xg, Ma:2011yi,
  Ahn:2012cg, Ishimori:2012fg, Ma:2012ez}, they have generally
considered the pattern of the vacuum expectation value (VEV) of the
triplet flavon to be $\propto (1, 1, 1)^T$.  Some earlier work has
separately considered other forms of the triplet flavon, such
as~\cite{Antusch:2011ic}.  In addition, recent literature has focused
on small perturbations from the $(1, 1, 1)^T$ triplet flavon structure
as a mechanism for generating a nonzero
$\theta_{13}$~\cite{King:2010bk, King:2011zj, Branco:2012vs,
  Callen:2012kd}, where this small perturbation may arise from a
vacuum misalignment correction.

In contrast, our work considers the full parameter space of $A_4$
flavons contributing to the Majorana mass matrix in the Type I seesaw.
Thus we consider triplet flavon vevs that are markedly different from
the usual TBM $(1, 1, 1)^T$ form simultaneously with the presence of
$1'$ and $1''$ flavons.  By looking at a completely general admixture
of possibly flavon vevs, we can definitively constrain the entire
$A_4$ parameter space in this Type I seesaw model.  We remark that for
the charged lepton masses and the Dirac neutrino masses, we introduce
the minimal field content to generate their mass matrices and only
introduce the full flavon field content for the RH neutrino masses.
Moreover, even though we consider a larger $A_4$ parameter space than
the earlier literature, we still retain predictivity, especially when
the triplet flavon vev pattern preserves a subgroup of $A_4$.

Here we concentrate on the group $A_{4}$, which is the smallest group
that contains a triplet representation.  Other groups that have been
utilized include $O(2)$ and $SO(2)$ symmetry~\cite{Heeck:2011aa},
$\Delta(3n^2)$ and $\Delta (6n^2)$~\cite{King:2009ap,
  Ferreira:2012ri}, $\Delta_{96}$~\cite{Ding:2012xx, King:2012in}, $Q_6$ with three
sterile neutrinos~\cite{Araki:2011zg}, and product groups of modular
$Z_n$ finite groups in various permutations~\cite{Ishimori:2012gv,
  Gupta:2011ct, He:2012yt, deAdelhartToorop:2011re, Toorop:2011jn,
  Ge:2011qn, Rashed:2011xe, He:2011kn, Aranda:2011rt, Lashin:2012gr}.
Other studies have focused on permutation symmetry $S_3$ and $S_4$
models~\cite{Meloni:2011fx, Zhou:2011nu, Meloni:2012ci, Dev:2012ns,
  Meloni:2012sy, Morisi:2011pm, Siyeon:2012zu}.  Other mixing
scenarios beyond TBM and related patterns include democratic mixing,
which has been studied in~\cite{Xing:2011at, Chao:2011sp,
  Araki:2011qy}, tetra-maximal mixing~\cite{Zhang:2011aw}, and
anarchic mixing~\cite{Hall:1999sn, Haba:2000be, deGouvea:2003xe,
  deGouvea:2012ac}.

The paper is organized as follows.  In~\secref{model}, we briefly
review the $A_4$ finite group symmetry and our type-I seesaw model
implementation.  In~\secref{numResult}, we present the results of our
parameter scan of the $A_{4}$ flavor vev space.  We conclude
in~\secref{conclusions}.  An intermediate step of our calculation is
presented in~\appref{generalvevsolns}.


\section{The $A_4$ model}
\label{sec:model}

We construct a Type I seesaw model based on an $A_4$ flavor
symmetry.  We include three right-handed neutrinos $N_i$, which
are Standard Model gauge singlets.  These neutrinos transform as a
triplet $\mathbf{3}$ under $A_4$.  We also assign the lepton $SU(2)$
doublet $L \sim \mathbf{3}$, charged lepton $SU(2)$ singlets $e_R \sim
\mathbf{1}$, $\mu_R \sim \mathbf{1^{\prime \prime}}$, $\tau_R \sim
\mathbf{1^\prime}$.  To separate the charged lepton coupling scalars from the neutrino coupling scalars, we impose a $Z_2$ symmetry.  These representations are summarized
in~\tableref{A4Reps}.

\begin{table}[t!]
\begin{tabular}{l|l|lllll|lllll}
& $H$ & $L$ & $N$ & 
$e_R$ & $\mu_R$ & $\tau_R$ & 
$\phi_E$ & $\phi_N$ & 
$\eta$ & $\chi$ & $\psi$ \\ 
\hline 
$A_4$ & 
\textbf{1} & \textbf{3} & \textbf{3} & 
\textbf{1} & $\mathbf{1^{\prime \prime}}$ & $\mathbf{1^\prime}$ & 
\textbf{3} & \textbf{3} & 
\textbf{1} & $\mathbf{1^\prime}$ & $\mathbf{1^{\prime \prime}}$ \\
$Z_2$ &
\textbf{1} & \textbf{1} & \textbf{1} & 
\textbf{-1} & \textbf{-1} & \textbf{-1} & 
\textbf{-1} & \textbf{1} & 
\textbf{1} & \textbf{1} & \textbf{1}
\end{tabular}
\caption{The $A_4$ and $Z_2$ charge assignments of the Standard Model fields and
  $A_4$ flavons.}
\label{table:A4Reps}
\end{table}

The Lagrangian for the leptons is
\begin{eqnarray}
\label{eqn:Lagrangian}
\mathcal{L} & \supset & \left( H \bar{L} 
\left( \lambda_e e_R + \lambda_\mu \mu_R + \lambda_\tau \tau_R \right) 
\left( \dfrac{\phi_E}{\Lambda} \right) +
\lambda_N \tilde{H} \bar{L} N + \text{ h.c.} \right)
\\
& & \quad 
+ \Lambda_{RR} N^T N 
\left( \dfrac{ c_N \phi_N + c_\eta \eta + c_\chi \chi + c_\psi \psi 
}{\Lambda} \right) + c.c. \ ,
\nonumber
\end{eqnarray}
where $\phi_E \sim \mathbf{3}$, $\phi_N \sim \mathbf{3}$, $\eta \sim
\mathbf{1}$, $\chi \sim \mathbf{1^\prime}$, and $\psi \sim
\mathbf{1^{\prime \prime}}$ are scalar fields which acquire vevs and
break the $A_4$ symmetry at the scale $\Lambda$, and the couplings
$c_N$, $c_\eta$, $c_\chi$, and $c_\psi$ are complex.  However, we absorb all phases (including Majorana phases) into $c_N$ and thus $c_N$ is indexed to be non-universal.  We will not specify
here the scalar potential to give the $\phi_E$, $\phi_N$, $\eta$,
$\chi$, and $\psi$ fields vevs, but instead we leave a study of the
potential construction and vacuum alignment questions for future work.
The Lagrangian includes the familiar Dirac masses for the charged
leptons, new Dirac masses for the neutrinos, and a general
$A_4$--invariant Majorana mass matrix which includes all possible
$A_4$ contractions.  Thus, in contrast with early, pre-Daya Bay and
RENO models that generated exact TBM mixing by including only the
$\phi_N \sim (1, 1, 1)^T$ and $\eta$ flavons, we include the $\chi$
and $\psi$ flavons and allow $\phi_N$ vev to be less constrained:
subcategories of our approach have also been considered
previously~\cite{King:2011zj, King:2011ab, Shimizu:2011xg, Ma:2011yi,
  Ahn:2012cg, Ishimori:2012fg, Ma:2012ez}.

The explicit form of the resulting mass matrices for the charged
leptons and neutrinos is easily obtained from the $A_4$ invariants,
which are reviewed, for example, in Ref.~\cite{Altarelli:2010gt}.  We
assume $\phi_E$ acquires a vev $\langle \phi_E \rangle = \Lambda
(1, 0, 0)^T$, and thus after electroweak symmetry breaking whereby the
Higgs acquires a vev $v_h$, the charged lepton mass matrix is
\begin{equation}
\label{eqn:charged_mass}
\bar{L} M_L e = \bar{L} v_h
\left( \begin{array}{ccc}
\lambda_e & 0 & 0 \\
0 & \lambda_\mu & 0 \\
0 & 0 & \lambda_\tau \\
\end{array} \right) e \ ,
\end{equation}
where $e = (e_R, \mu_R, \tau_R)$.  The corresponding Dirac mass matrix
for the neutrinos is simply governed by the $A_4$ contraction of $L$
and $N$, giving
\begin{equation}
\label{eqn:Dirac_mass}
\lambda_N \tilde{H} \bar{L} N = \bar{L} M_{Dc} N
= \bar{L}
\left( \begin{array}{ccc}
\lambda_N v_h & 0 & 0 \\
0 & 0 & \lambda_N v_h \\
0 & \lambda_N v_h & 0 \\
\end{array} \right) N \ .
\end{equation}

For the Majorana mass matrix, we will be more general and allow
$\phi_N$ to obtain a general vev pattern, $c_N \langle \phi_N
\rangle = \Lambda (\phi_a, \phi_b, \phi_c)$. 
In addition, we let
$c_\eta \langle \eta \rangle = \Lambda \eta$, and similarly for
$\chi$ and $\psi$, such that the vev parameters $\eta$, $\chi$, and
$\psi$ are dimensionless and have absorbed their respective Lagrangian
couplings.  We have the Majorana mass matrix
\begin{equation}
\label{eqn:Majorana_mass}
M_N N^T N = \Lambda_{RR} N^T N \left( \begin{array}{ccc}
\frac{2}{3} \phi_a + \eta & -\frac{1}{3} \phi_c + \psi 
 & -\frac{1}{3} \phi_b + \chi \\ 
-\frac{1}{3} \phi_c + \psi & \frac{2}{3} \phi_b + \chi
 & -\frac{1}{3} \phi_a + \eta \\ 
-\frac{1}{3} \phi_b + \chi & -\frac{1}{3} \phi_a + \eta 
 & \frac{2}{3} \phi_c + \psi \\
\end{array} \right) \ .
\end{equation}

It is worth noting that there is another term one could write, $m_{RR} N^T N$, which sets an additional mass scale $m_{RR}$ which we will designate as  $\Lambda_{RR}$.  We absorb this term into the vev of $\eta$.  Since $N$ is a gauge singlet, there is no connection between the Higgs
vev and the mass scale $\Lambda_{RR}$.  In particular, if we set
$\Lambda_{RR} \sim \mathcal{O} (\Lambda_{GUT}) \approx 10^{16}$ GeV,
we exercise the seesaw mechanism to generate light neutrino masses.
Moreover, we assume the $A_4$ breaking scale $\Lambda \sim 0.1
\Lambda_{RR}$ to avoid tuning issues between the mass scale and the breaking scale.  The block matrix for the neutrinos in the $(\nu_L,
N)^T$ basis is
\begin{equation}
M_{\nu} = \left( \begin{array}{cc}
0 & M_{Dc}^T \\
M_{Dc} & M_N \\
\end{array} \right) \ ,
\end{equation}
which generates the effective neutrino mass matrix
\begin{equation}
\label{eqn:Mnu_eff1}
M_{\nu, \text{ eff}} = M_{Dc} M_N^{-1} M_{Dc}^T \ ,
\end{equation}
after the right-handed neutrinos have been integrated out.

Now, in exact analogy with the Cabibbo-Kobayashi-Maskawa (CKM) matrix,
the PMNS matrix arises when we express the weak charged current
interactions in the lepton mass basis.  The PMNS matrix is
\begin{equation}
\label{eqn:PMNS}
V_{PMNS} = U_L U_\nu^\dagger \ ,
\end{equation}
where
\begin{equation}
\label{eqn:UL}
M_L^{\text{diag}} = U_L M_L U_L^\dagger \ , \quad
M_\nu^{\text{diag}} = U_\nu M_{\nu, \text{ eff}} U_\nu^\dagger \ ,
\end{equation}
but from~\eqnref{charged_mass}, $U_L = \mathbf{1}_3$, and thus the
PMNS matrix is identified with the neutrino diagonalization matrix
$U_\nu^\dagger$.  The three parameters $\lambda_e$, $\lambda_\mu$, and
$\lambda_\tau$, which are each rescaled by $\langle \phi_E \rangle /
\Lambda$, are in one-to-one correspondence with the three charged
lepton masses and are hence fixed.

Having identified $V_{PMNS} \equiv U_\nu^\dagger$, we adopt the
standard parametrization of the PMNS matrix given by
\begin{equation}
\label{eqn:VPMNS}
V_{PMNS} = \left( \begin{array}{ccc} 
1 & 0 & 0 \\
0 & c_{23} & s_{23} \\
0 & -s_{23} & c_{23} \\
\end{array} \right)
\left( \begin{array}{ccc}
c_{13} & 0 & s_{13} e^{-i \delta} \\
0 & 1 & 0 \\
-s_{13} e^{i \delta} & 0 & c_{13} \\
\end{array} \right)
\left( \begin{array}{ccc}
c_{12} & s_{12} & 0 \\
-s_{12} & c_{12} & 0 \\
0 & 0 & 1 \\
\end{array} \right)
\text{ diag}(1, e^{i \xi_1}, e^{i \xi_2} ) \ ,
\end{equation}
where $c_{ij} = \cos \theta_{ij}$, $s_{ij} = \sin \theta_{ij}$,
$\delta$ is the Dirac phase, $\xi_1$ and $\xi_2$ are the
Majorana phases, with $0 \leq \theta_{ij} \leq \pi /2$, $0 \leq
\delta$, $\xi_1,\xi_2 \leq 2 \pi$.

We now have a solvable system of equations relating the neutrino
masses, mixing angles, and phases with the flavon vevs.
From~\eqnref{Mnu_eff1},~\eqnref{UL}, and using $U_\nu^\dagger =
V_{PMNS}$, we find the relation
\begin{equation}
\label{eqn:Mnu_eff}
M_{Dc} M_N^{-1} M_{Dc}^T = M_{\nu \text{ eff}} = 
U_\nu^\dagger M_\nu^{\text{diag}} U_\nu
= V_{PMNS} M_\nu^{\text{diag}} V_{PMNS}^\dagger \ ,
\end{equation}
where $M_N$ is given in~\eqnref{Majorana_mass}.  Solving for
$M_N^{-1}$ and inverting, we get
\begin{equation}
\label{eqn:MN_system}
M_N = M_{Dc}^T V_{PMNS} (M_\nu^{\text{diag}})^{-1} V_{PMNS}^\dagger M_{Dc} \ .
\end{equation}
Entry by entry, we have a system of six equations that can be solved
analytically, which gives exact relations between the $A_4$ parameter
space and the physical masses and PMNS mixing angles.  This is
manifestly symmetric, which is most easily seen
from~\eqnref{Majorana_mass}.  A similar result using the mass entries
of the Majorana mass matrix instead of the triplet and one-dimensional
flavon vevs was the starting point of~\cite{Ishimori:2012fg}.  Our
approach, in contrast, directly shows the correlation between
different UV triplet flavons and low energy neutrino observables.

Naively, we have a six-dimensional UV parameter space, given by the
three components of the $\phi_N$ triplet flavon and each of the
one-dimensional flavons.  The flavon breaking scale $\Lambda$ and the
Majorana mass scale are unobservable and can be absorbed into the six
flavon vev components.  Now, although the triplet's vevs can be
independent degrees of freedom, certain breaking alignments in flavon
space preserve the $Z_2$ or $Z_3$ subgroups of $A_4$ and thus reduce
the number of UV parameters.  We will thus categorize our results
according to subgroup preserving and subgroup breaking triplet vev
patterns, which are listed in~\tableref{A4vevs}.  Subgroup preserving
vev patterns thus effectively have a four-dimensional parameter space,
while subgroup breaking vev patterns have a six-dimensional space.

In general, there are nine physical parameters which give rise to
eight physical predictions: three mixing angles, three masses, and one
Dirac $CP$ phase, as well as two Majorana phases that combine to
dictate the rate of neutrinoless double $\beta$ decay.  We will not
discuss $0\nu 2\beta$ further in this paper, and instead focus on the
three angles, three masses, and the Dirac $CP$ phase.  We can thus see
both our subgroup preserving and subgroup breaking categories are
predictive: the four parameters in the subgroup preserving category,
for instance, are over-constrained by the existing neutrino
measurements, and the existence of a nontrivial solution reflects the
suitability of the $A_4$ finite group as a possible flavor symmetry of
the lepton sector.  In addition, for both categories, the yet-to-be-discovered 
Dirac $CP$ phase is predicted from our parameter scan, which we detail
in the next section.
\begin{table}[t!]
\begin{tabular}{{|c|c|}}
\hline
Subgroup & vev alignment \\ \hline \hline
$Z_2$ & $(1, 0, 0), \, (0, 1, 0), \, (0, 0, 1)$ \\ \hline
$Z_3$ & $(-1, 1, 1), \, (1, 1, -1), \, (1, -1, 1), \, (1, 1, 1)$ \\ \hline
breaking & 
$(0, 1, 1), \, (1, 0, 1), \, (1, 1, 0), \, (0, 1, -1), \, 
 (2, 1, 1), \, (1, 1, 2), \, (1, 2, 1), \, (1, -2, 1), \, 
 (1, 1, -2), \, (-2, 1, 1)$\\ \hline
\end{tabular}
\caption{Listed vev alignments for $\phi$ that preserve a $Z_2$ or
  $Z_3$ subgroup of $A_4$.}
\label{table:A4vevs}
\end{table}

\subsection{Breaking Bimaximality, Analytic Results}
\label{subsec:bimaximal_breaking}
We present an analytic understanding of breaking bimaximality ({\it
  i.e.} deviations of $\theta_{23}$ from $45^\circ$) in our general
Type I seesaw $A_4$ construction.  First, we calculate the effective
neutrino mass matrix, and then analyze to the extent a $\theta_{23} =
45^\circ$ rotation diagonalizes this matrix.  Starting
with~\eqnref{Dirac_mass}, we write
\begin{equation}
M_{Dc} = M_{Dc}^T = \lambda_N v_h \left( \begin{array}{ccc}
1 & 0 & 0 \\
0 & 0 & 1 \\
0 & 1 & 0 \\
\end{array} \right) \ ,
\end{equation}
and we can recognize that $M_{Dc}^{-1} = \dfrac{1}{\lambda_N^2 v_h^2}
M_{Dc}$.  Next, since $U_\nu$ is the diagonalization matrix of
$M_{\nu, \text{ eff}}$, as established in~\eqnref{UL}, $U_\nu$ is also
the diagonalization matrix of $M_{\nu, \text{ eff}}^{-1}$, since
\begin{equation}
(M_{\nu}^{\text{diag}})^{-1} = (U_\nu M_{\nu, \text{ eff}} U_\nu^\dagger)^{-1}
= (U_\nu^\dagger)^{-1} (M_{\nu, \text{ eff}})^{-1} (U_\nu)^{-1}
= U_\nu (M_{\nu, \text{ eff}})^{-1} U_\nu^\dagger \ .
\end{equation}
Thus, the seesaw mass matrix in~\eqnref{Mnu_eff1} can be inverted to
give
\begin{equation}
\label{eqn:Mnu_effinverse}
M_{\nu, \text{ eff}}^{-1} = \frac{1}{\lambda_N^4 v_h^4} M_{Dc} M_N M_{Dc}^T \ .
\end{equation}
Recall that because our charged lepton mass matrix
in~\eqnref{charged_mass} is diagonal, we have the identity $V_{PMNS} =
U_{\nu}^\dagger$, so the diagonalization matrix of $M_{\nu, \text{
    eff}}$, and by extension, $M_{\nu, \text{ eff}}^{-1}$ is the PMNS
matrix.

Recall that the PMNS matrix is composed of three Jacobi rotation
angles, ordered as a $\theta_{23}$ rotation, a $\theta_{13}$ rotation,
and then a $\theta_{12}$ rotation, and the phase rotations, as we can
see from~\eqnref{VPMNS}.  Also, recall that a Jacobi rotation operates
on a $2 \times 2$ block of the matrix, such as
\begin{equation}
\left( \begin{array}{cc}
a & b \\
b & c \\
\end{array} \right) \ ,
\end{equation}
where the rotation angle is defined as
\begin{equation}
\tan 2 \theta = \dfrac{2b}{c - a} \ .
\end{equation}
In particular, for fixed $c$ and $a$, the sign of $b$ determines the
sign of $\theta$.  Moreover, we know that a rotation matrix for a
single angle $\alpha + \beta$ can be decomposed into first rotating by
$\alpha$ and then by $\beta$.  (This, of course, does not commute with
rotations about other axes.)  Hence, we can rotate $M_{\nu, \text{
    eff}}^{-1}$ by $\theta_{23} = 45^\circ$ and then understand
deviations from bimaximality by testing the remaining presence of
off-diagonal entries in the (3, 2) and (2, 3) entries of the effective
neutrino mass matrix.

From~\eqnref{Dirac_mass} and~\eqnref{Majorana_mass}, the RHS
of~\eqnref{Mnu_effinverse} is
\begin{equation}
\label{eqn:Mnu_effinverseRHS}
M_{\nu, \text{ eff}}^{-1} = \frac{1}{\lambda_N^2 v_h^2}
\left( \begin{array}{ccc}
\frac{2}{3} \phi_a + \eta & -\frac{1}{3} \phi_b + \chi & 
-\frac{1}{3} \phi_c + \psi \\
-\frac{1}{3} \phi_b + \chi & \frac{2}{3} \phi_c + \psi &
-\frac{1}{3} \phi_a + \eta \\
-\frac{1}{3} \phi_c + \psi & -\frac{1}{3} \phi_a + \eta &
\frac{2}{3} \phi_b + \chi \\
\end{array} \right) \ ,
\end{equation}
which as a trivial check, is still symmetric.  After we perform the
$\theta_{23} = 45^\circ$ bimaximal rotation, we have
\begin{equation}
\begin{array}{ll}
&R_{BM}^T M_{\nu, \text{ eff}}^{-1} R_{BM} = \\
&\left( \begin{array}{ccc}
\frac{2}{3} \phi_a + \eta & 
\frac{1}{3 \sqrt{2}} (-\phi_b + \phi_c) + \frac{1}{\sqrt{2}} (\chi - \psi) 
& \frac{-1}{3 \sqrt{2}} (\phi_b + \phi_c) + \frac{1}{\sqrt{2}} (\chi + \psi) \\
\frac{1}{3 \sqrt{2}} (-\phi_b + \phi_c) + \frac{1}{\sqrt{2}} (\chi - \psi) &
\frac{1}{3} (\phi_a + \phi_b + \phi_c) - \eta + \frac{1}{2} \chi + \frac{1}{2} \psi & 
\frac{1}{3} (-\phi_b + \phi_c) - \frac{1}{2} \chi + \frac{1}{2} \psi \\
\frac{-1}{3 \sqrt{2}} (\phi_b + \phi_c) + \frac{1}{\sqrt{2}} (\chi + \psi) &
\frac{1}{3} (-\phi_b + \phi_c) - \frac{1}{2} \chi + \frac{1}{2} \psi &
\frac{1}{3} (-\phi_a + \phi_b + \phi_c) + \eta + \frac{1}{2} \chi + \frac{1}{2} \psi \\
\end{array} \right) .
\end{array}
\end{equation}
So, if $\frac{1}{3} (-\phi_b + \phi_c) - \frac{1}{2} \chi +
\frac{1}{2} \psi \neq 0$, then the $\theta_{23} = 45^\circ$ bimaximal
rotation was insufficient to eliminate the (3, 2) and (2, 3) entries
and an additional $\theta_{23}$ rotation is needed.  (In addition, if
the (2, 2) and (3, 3) entries are identical, then $\theta_{23} =
45^\circ$ is guaranteed.)  Moreover, we see that the transformation
$\phi_b \leftrightarrow \phi_c$ and $\chi \leftrightarrow \psi$
changes the sign of the (3, 2) and (2, 3) entries while leaving the
(2, 2) and (3, 3) entries fixed.  Hence, we can see that any viable
solution characterized by a triplet vev of $(\phi_a, \phi_b, \phi_c)$
and a particular set of $\eta$, $\chi$, and $\psi$ can be transmuted
to a different solution characterized by $(\phi_a, \phi_c, \phi_b)$
and $\eta$, $\psi$, and $\chi$ with an opposite sign of the deviation
from $\theta_{23} = 45^\circ$.


\section{Parameter Scan Results for Subgroup Preserving and 
Subgroup Breaking Triplet Flavon VEVs}
\label{sec:numResult}

Since current neutrino experiments do not have sensitivity to
individual neutrino masses, we constrain the low energy neutrino
observables by fitting to three mixing angles, two $\Delta m^2$, 
and the cosmological constraint on the sum of absolute 
neutrino masses.  Clearly, the remaining parameter space for $\delta$
is the predictive relation from our parameter scan.  (As stated
before, we do not discuss the sensitivity to the Majorana phases from
neutrino-less double $\beta$ decay experiments.)  We first take the
system of equations in~\eqnref{MN_system} and solve for the $A_4$
breaking vevs in terms of the neutrino observables. These solutions
are presented in~\appref{generalvevsolns}.  Since we want to work from
the top-down, however, we partially invert the system to solve for the
neutrino masses and the $A_4$ singlet vevs in terms of the
$A_4$ triplet vev and the neutrino mixing angles and the $CP$ phase.  We
obtain
\begin{equation}
m_i=\frac{v_h^2}{\Lambda_{RR}} (a^i b^j c^k \epsilon_{ijk})
(\phi_a b^k c^j \epsilon_{ijk} + \phi_b a^j c^k \epsilon_{ijk} +
\phi_c a^k b^j \epsilon_{ijk})^{-1} \ ,
\label{eqn:numass}
\end{equation}
where $m_1, m_2$, $m_3$ are the three light neutrino masses, $i$, $j$,
$k$ = 1, 2, 3, and $\epsilon_{ijk}$ is the Levi-Civita tensor. The
three-component vectors $\vec{a}$, $\vec{b}$, $\vec{c}$ are found
in~\tableref{inversionCoeff}.

\begin{table}[t!]
\begin{tabular}{|l|l|}
\hline
$a_1$ & 
$ \left[ 
c^2_{12} c^2_{13} - \left( s_{12} s_{23} - e^{-i\delta} c_{12} c_{23} s_{13} \right) 
\left( -c_{23} s_{12} - e^{i\delta} c_{12} s_{13} s_{23} \right) 
\right]$ \\
\hline
$a_2$ &
$e^{-i 2\xi_1} \left[ 
c^2_{13} s^2_{12} + \left( c_{12} s_{23} + e^{-i\delta} c_{23} s_{12} s_{13} \right)
\left( c_{12} c_{23} - e^{i\delta} s_{12} s_{13} s_{23} \right) 
\right]$ \\ \hline
$a_3$ &
$e^{-i 2\xi_2}(s^2_{13} - c^2_{13} c_{23} s_{23})$ \\ \hline
$b_1$ & 
$ \left[ 
\left( s_{12} s_{23} - e^{-i\delta} c_{12} c_{23} s_{13} \right)
\left( s_{12} s_{23} - e^{ i\delta} c_{12} c_{23} s_{13} \right)
+ c_{12} c_{13} \left( c_{23} s_{12} + e^{-i\delta} c_{12} s_{13} s_{23} \right) 
\right] $\\ \hline
$b_2$ & 
$e^{-i 2\xi_1} \left[
\left( c_{12} s_{23} + e^{-i\delta} c_{23} s_{12} s_{13} \right)
\left( c_{12} s_{23} + e^{ i\delta} c_{23} s_{12} s_{13} \right)
- c_{13} s_{12} \left( c_{12} c_{23} - e^{-i\delta} s_{12} s_{13} s_{23} \right)
\right] $\\ \hline
$b_3$ &
$e^{-i 2\xi_2}(c^2_{13} c^2_{23} - e^{-i\delta} c_{13} s_{13} s_{23})$ \\ \hline
$c_1$ &
$ \left[
c_{12} c_{13} \left( -s_{12} s_{23} + e^{-i\delta} c_{12} c_{23} s_{13} \right) 
+ \left( c_{23} s_{12} + e^{ i\delta} c_{12} s_{13} s_{23} \right) 
  \left( c_{23} s_{12} + e^{-i\delta} c_{12} s_{13} s_{23} \right) 
\right]$ \\ \hline
$c_2$ & 
$e^{-i 2\xi_1} \left[ 
c_{13} s_{12} \left( c_{12} s_{23} + e^{-i\delta} c_{23} s_{12} s_{13} \right)
+ \left( c_{12} c_{23} - e^{ i\delta} s_{12} s_{13} s_{23} \right)
  \left( c_{12} c_{23} - e^{-i\delta} s_{12} s_{13} s_{23} \right)
\right]$ \\ \hline
$c_3$ &
$e^{-i 2\xi_2}(c^2_{13} s^2_{23} - e^{-i\delta} c_{13} c_{23} s_{13})$ \\ \hline
\end{tabular}
\label{table:inversionCoeff}
\caption{The explicit components of $\vec{a}$, $\vec{b}$, $\vec{c}$ as
  function of mixing angles and the triplet $\phi$.}
\end{table} 

This partial inversion is advantageous because, using the measured
mixing angles, we can test individual $A_4$ triplet vevs and determine
the fit to the correct mass squared differences and the cosmological
constraint.  We reparametrize the mass squared difference constraints
into a ratio $\Delta_{21} / \Delta_{31}$, where $\Delta_{ij} \equiv
m_i^2 - m_j^2$ and use the dimensionless ratio to constrain the
dimensionless vevs in~\eqnref{numass}.  The mass scale $v_h^2 /
\Lambda_{RR}$ is then fixed by matching either of the measured mass
squared differences. 

In line with our top-down approach and intuition about the $A_4$
finite group, we attempt to preserve many of the symmetries present in
the case of TBM mixing.  Yet, simply relaxing $\sin^2 \theta_{13} = 0$
while simultaneously keeping the other TBM mixing angle relations did
not lead to viable solutions without perturbing the triplet vev
alignment. We therefore choose to relax the bimaximal relation $\sin^2
\theta_{23} = \frac{1}{2}$ and maintain the trimaximal relation
$\sin^2 \theta_{12} = \frac{1}{3}$.  In this way, we can understand a
larger region of $A_4$ parameter space since the experimental bounds
on $\theta_{12}$ are tighter than those on $\theta_{23}$. Given this
fixed $\theta_{12}$, we then choose each of the Dirac and Majorana
phases to be $0$ or $\pi$ for each triplet vev pattern.

However, this process itself is not entirely trivial, as not all
choices are independent. Up to field re-phasing, there are four
transformations about bimaximal mixing in which components
of~\eqnref{numass} transform antisymmetrically. Under the interchange
of $\theta_{23}=45^\circ+x\leftrightarrow\theta_{23}=45^\circ-x$ and
any of the four,
\begin{align}
\delta = 0, \xi_1 = 0, \xi_2 = 0 & \leftrightarrow 
\delta = \pi, \xi_1 = \pi, \xi_2 = \pi \\ 
\nonumber
\delta = 0, \xi_1 = \pi, \xi_2 = 0 & 
\leftrightarrow \delta = \pi, \xi_1 = \pi, \xi_2 = 0 \\
\nonumber
\delta = 0, \xi_1 = 0, \xi_2 = \pi & 
\leftrightarrow 
\delta = \pi, \xi_1 = 0, \xi_2 = \pi \\
\nonumber
\delta = 0, \xi_1 = \pi, \xi_2 = \pi &
\leftrightarrow \delta = \pi, \xi_1 = 0, \xi_2 = 0 \ , \nonumber
\end{align}
we find
\begin{align}
 \vec{a} \cdot (\vec{b} \times \vec{c}) & \leftrightarrow 
-\vec{a} \cdot (\vec{b} \times \vec{c}) \\ 
\nonumber
\vec{b} \times \vec{c} & \leftrightarrow \vec{c} \times \vec{b} \\ 
\nonumber
\vec{a} \times \vec{b} & \leftrightarrow \vec{a} \times \vec{c} \ .
\end{align}

It immediately follows that the existence of one solution implies
there is a corresponding solution with different phases (and a
possibly different triplet alignment) which is related under these
transformations.  What remains to be specified are the angles
$\theta_{23}$ and $\theta_{13}$, which we scan over the $2 \sigma$
$\theta_{23}$ range of~\cite{Schwetz:2011qt} and over the $2 \sigma$
$\theta_{13}$ experimental bounds from Daya Bay in
Ref.~\cite{An:2012eh} . We draw contours in this two-dimensional plane
that satisfy the measured mass squared differences
from~\cite{Schwetz:2011qt} and we constrain the ratio of $\Delta m^2$
within $2\sigma$ uncertainties as dictated by the predicted neutrino
hierarchy and set the individual neutrino mass scale using
$\Lambda_{RR}$.  From the existence of matter effects in solar
neutrino oscillation, we know that $\Delta_{21}>0$. So by enforcing
this constrain upon the ratio there is a unique positive value and
negative value for $\Delta_{21}/\Delta_{31}$ that correspond to a
normal or inverted neutrino mass hierarchy respectively.  We then
require the contours to satisfy the $\sum\limits_i |m_{\nu_i}| < 0.81$
eV~\cite{Beringer:2012}.  Because degenerate neutrino spectra lead to
numerically unstable results, we discard possible solutions arising
from degenerate neutrino spectra.

\subsection{Numerical Results}

\begin{figure}[t!]
\includegraphics[width=0.48\columnwidth]{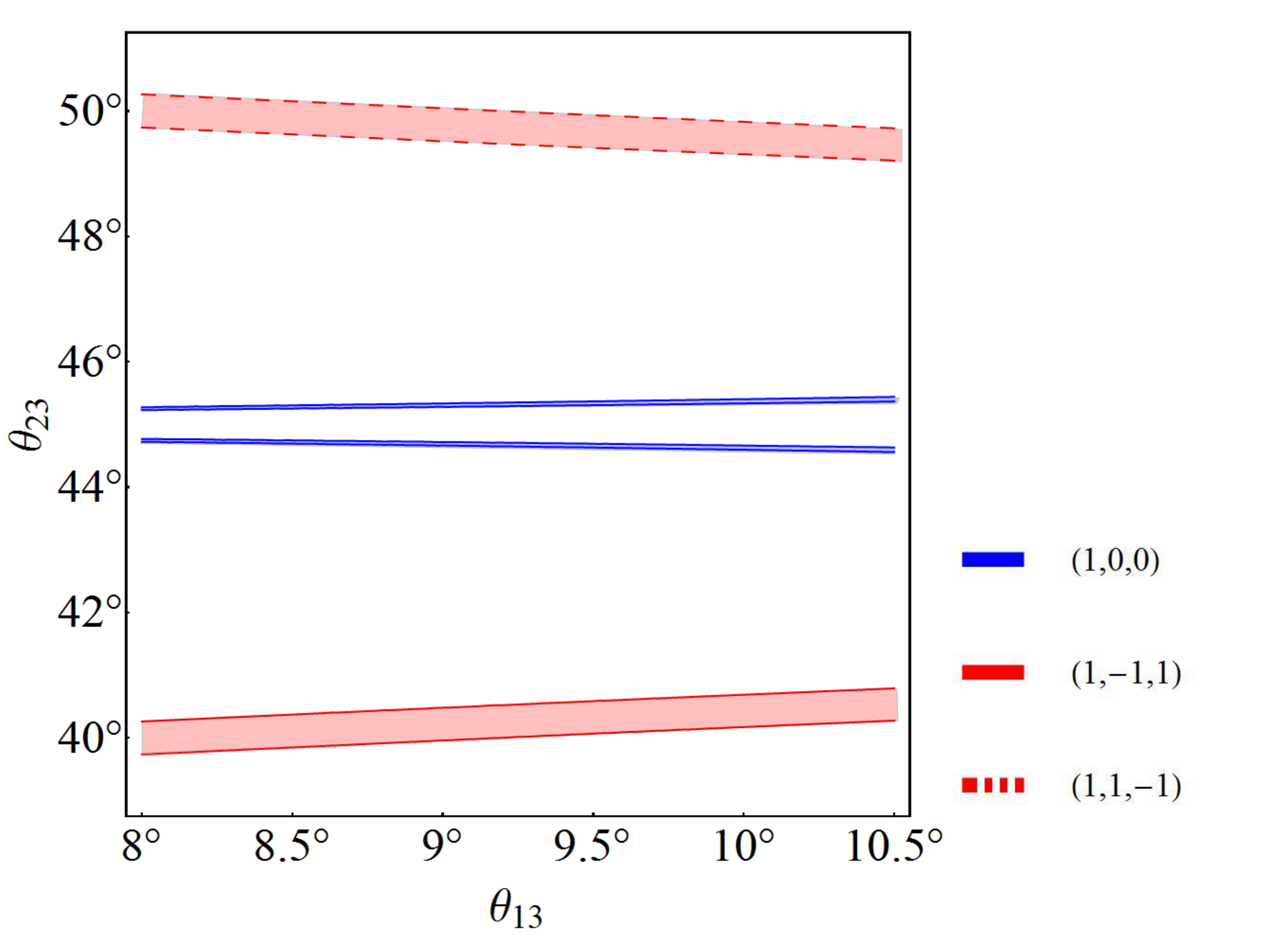} 
\includegraphics[width=0.48\columnwidth]{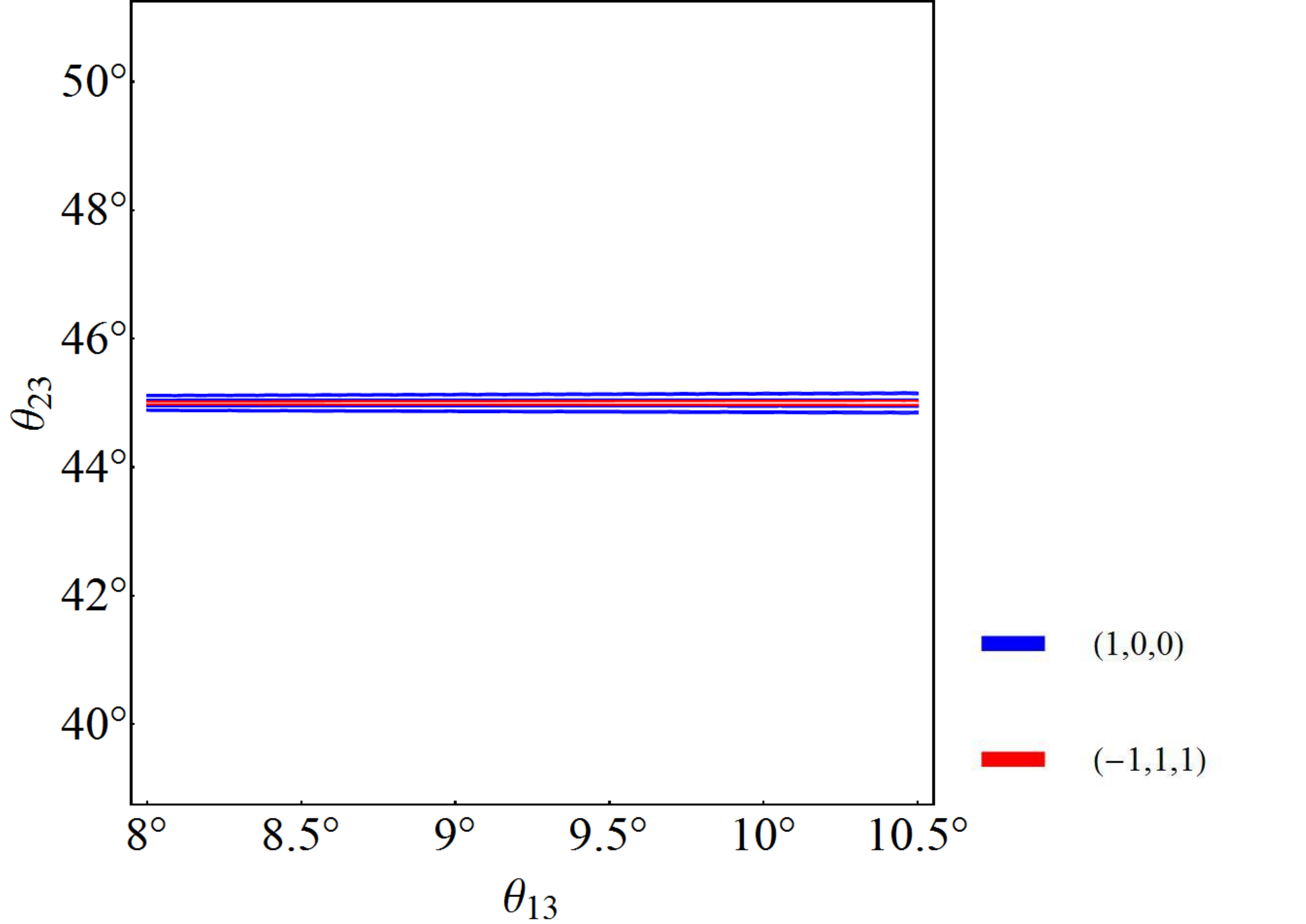} 
\caption{Contour plots of $\Delta_{21}/\Delta_{31}$
  over the 2$\sigma$ regions of $\theta_{23}$ and $\theta_{13}$ for
  subgroup preserving triplet vev patterns that generate (left panel) a
  normal hierarchy and (right panel) an inverted hierarchy.  For normal
  hierarchies, we find the $(1, 0, 0)$ pattern stays close to the
  central angle of $\theta_{23}$, while the the $(1, -1, 1)$ and $(1,
  1, -1)$ patterns avoid the central value.  The inverted hierarchy
  solutions, however, only deviate from the maximal angle of
  $\theta_{23}$ by values of order $\mathcal{O}(0.1^\circ)$.}
\label{fig:spni}
\end{figure}

These results are shown in the left panel of~\figref{spni} for
subgroup preserving triplet vev patterns from~\tableref{A4vevs} that
generate a normal hierarchy, while the right panel shows the subgroup
preserving patterns that generate an inverted hierarchy.  These
figures clearly show the correlations between the deviation from the
bimaximal $\theta_{23}$ and nonzero $\theta_{13}$.  Patterns that have
such solutions are now uniquely determined and predict a very specific
combination of masses and the $CP$ phase. A notably absent alignment
is the (1,1,1) breaking pattern, often associated with TBM mixing.  We
can understand this by considering~\eqnref{Mnu_effinverse} and
asserting $\phi_a = \phi_b = \phi_c$.  We find
\begin{equation}
f(\theta_{12}, \theta_{13}, \theta_{23}) (m_2 - m_3) = 0 \ ,
\end{equation}
and so unless the function $f$ is zero, the masses $m_2$ and $m_3$ are
required to be degenerate and this vev is not a viable
phenomenological alignment. Under exact TBM mixing, $f$ is zero and
the masses are free.

\begin{figure}[t!]
\includegraphics[width=0.48\columnwidth]{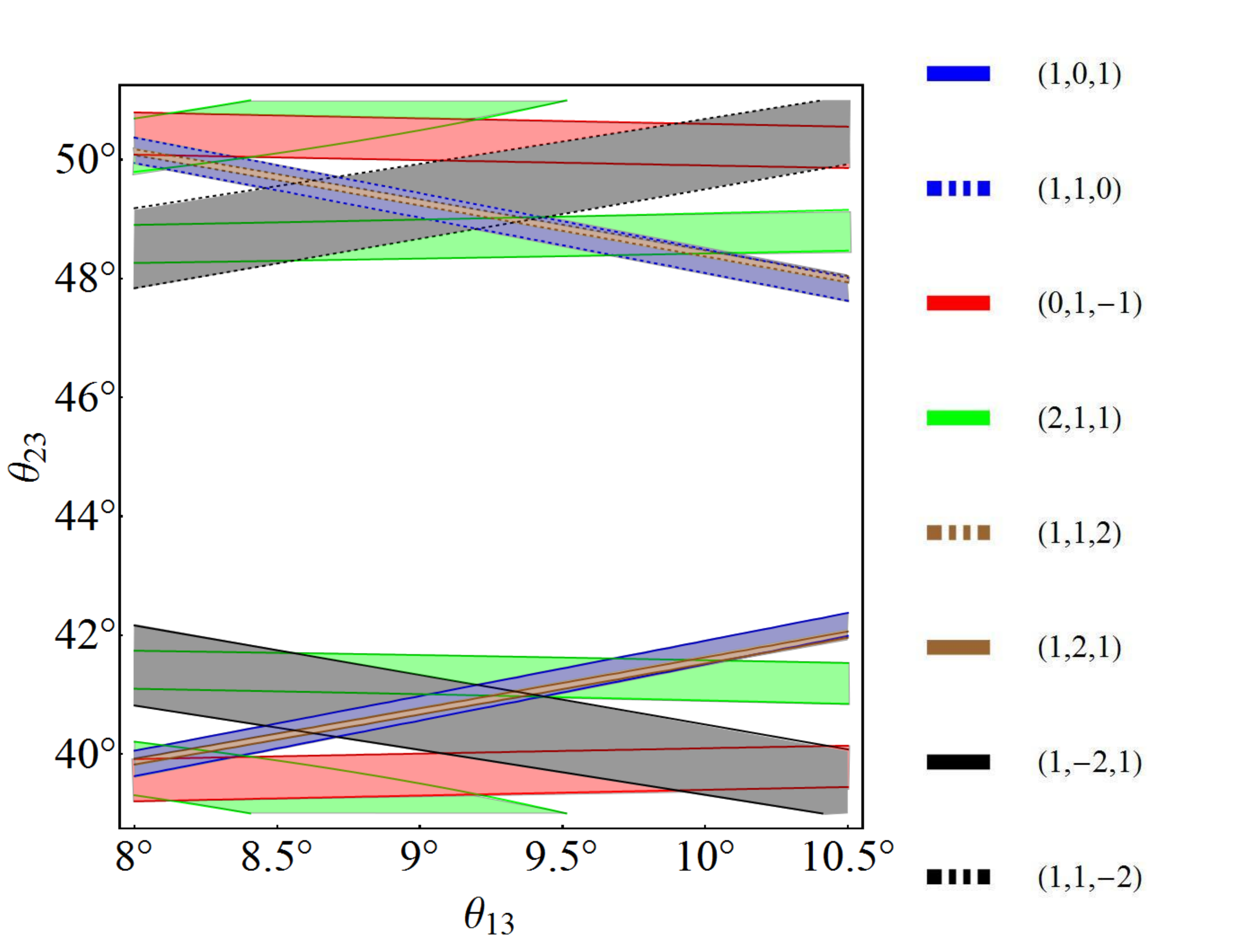} 
\includegraphics[width=0.48\columnwidth]{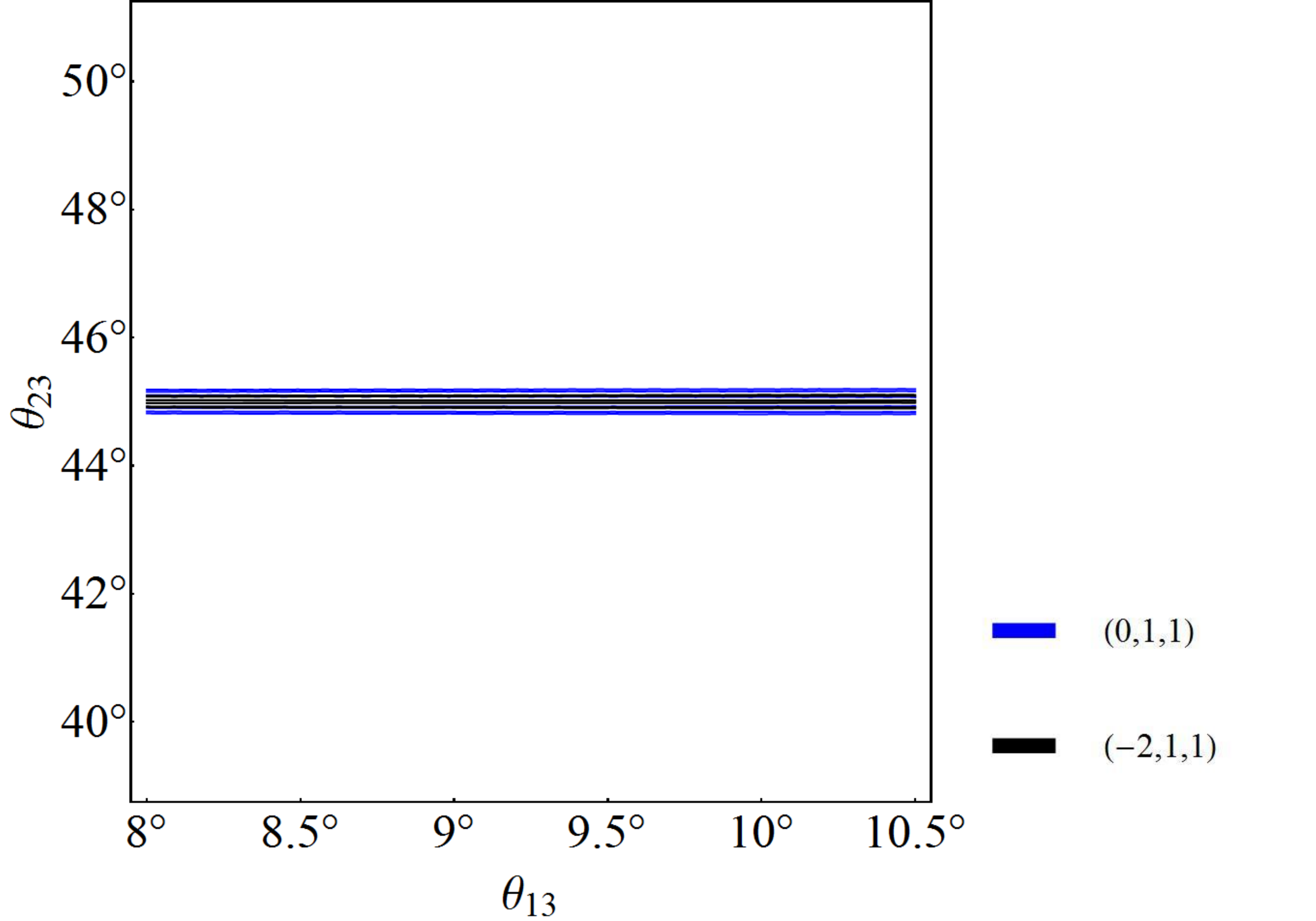} 
\caption{Contour plots of $\Delta_{21}/\Delta_{31}$
  over the 2$\sigma$ regions of $\theta_{23}$ and $\theta_{13}$ for
  subgroup breaking triplet vev patterns that generate (left panel)
  a normal hierarchy and (right panel) an inverted hierarchy.  The
  normal hierarchy solutions characteristically avoid the central
  value of $\theta_{23}$.  Like the subgroup preserving vev patterns,
  here the inverted hierarchy solutions feature little deviations from
  the central value of $\theta_{23}$}
\label{fig:soni}
\end{figure}

We also show the corresponding parameter scans for subgroup breaking
triplet vev patterns that give normal hierarchies in the left panel
of~\figref{soni} and inverted hierarchies in the right panel.
Although the parameter space is less predictive, we can nevertheless
see that a normal hierarchy requires significant breaking of
bimaximality, while an inverted hierarchy still allows for a bimaximal
$\theta_{23}$ for some vev patterns.  

For each successful triplet vev pattern and phase choice, we list the
$\theta_{23}$ value and neutrino mass spectrum corresponding to the
central value for $\theta_{13}$ from the Daya Bay
collaboration~\cite{An:2012eh} in~\tableref{vevlist}.

The vev alignments $(1, 0, 0)$ and the permutations of $(-1, 1, 1)$
comprising the first and second sets of solutions shown
in~\tableref{vevlist} are vevs preserving the $Z_3$ and $Z_2$
subgroups, respectively.  The remaining vev alignments break $A_4$
completely.  Each vev alignment can produce some deviation from
$\theta_{23} = \pi/4$, and most notably, non-$Z_3$ preserving
alignments can produce significant breaking of bimaximality where TBM
cannot be approximate.

Generally, we find that inverted hierarchies feature deviations from
$\theta_{23}$ from the TBM case of order $\mathcal{O}(0.1^\circ)$,
while normal hierarchies favor deviations an order of magnitude
larger. For a given choice of Dirac $CP$ phase, we find that most
breaking patterns are relatively insensitive to changes in
$\theta_{13}$ with the current experimental bounds.  We also highlight
the fact that many simple $A_4$ triplet vev possibilities are excluded
by the current experimental data.  In particular, arbitrary triplet
vev patterns in general will not generate the small hierarchy in
experimental mass squared differences, and those that do are highly
predictive about the extent of bimaximal breaking.

We can semi-analytically see why the normal hierarchy solutions
deviate more strongly from bimaximality than the inverted hierarchy
solutions.  In general, for fixed $\theta_{12}$ and $\theta_{13}$, we
can consider a Taylor expansion of the neutrino masses around $z
\equiv \theta_{23} - 45^\circ$,
\begin{align}
m_1 &= x_1 + y_1 z + \ldots \ , \\
m_2 &= x_2 + y_2 z + \ldots \ , \nonumber \\
m_3 &= x_3 + y_3 z + \ldots \ , \nonumber
\end{align}
where $x_{1,2,3}$ and $y_{1,2,3}$ are the zeroth and first order
coefficients.  Using this expansion, we have
\begin{equation}
\Delta_{21} = m_2^2 - m_1^2 \approx 
x_2^2 - x_1^2 + 2 (x_2 y_2 - x_1 y_1) z + \mathcal{O} (z^2) \ .
\end{equation}
A complete expression of this mass squared difference for the central
values of $\theta_{12}$ and $\theta_{13}$ and arbitrary triplet vev is
not useful, but if we consider the special case of $\phi_b = \phi_c$, we
generally find $x_1 = x_2$ and $y_1 = -y_2$, giving
\begin{equation}
\Delta_{21} \approx 4 x_2 y_2 z + \mathcal{O} (z^2) \ .
\end{equation}
Furthermore, if we consider the behavior of the expansion coefficients
$x_1$ and $y_1$ for normal hierarchy solutions versus related inverted
hierarchy solutions, we find $x_1^{(N)} \approx x_1^{(I)}$ and $4
y_1^{(N)} \approx y_1^{(I)}$: from the $\Delta_{21}$ constraint, we
see
\begin{equation}
z^{(N)} \approx 4 z^{(I)} \ ,
\end{equation}
which implies that the resulting deviation from bimaximality is much
larger for normal hierarchies compared to inverted hierarchies.  When
exact expressions are used, the difference can be as big as an order
of magnitude, as evident in the figures.

\begin{table}[b!]
\begin{tabular}{|c|c|c|c|c|}
\hline
$(\phi_a^{\nu}, \phi_b^{\nu}, \phi_c^{\nu})$ & mass hierarchy & 
$(\xi_1, \xi_2, \delta, \theta_{23} - 45^{\circ})$ & $(m_1, m_2, m_3)$ \\
\hline
$(1,0,0)$ & N & $(0, 0, \pi, -0.3^{\circ}); \, (\pi, \pi, 0, 0.3^{\circ})$ & $(0.0447, 0.0455, -0.0667)$ \\
$(1,0,0)$ & I & $(\pi, 0, 0, -0.05^{\circ}); \, (\pi, 0, \pi, 0.05^{\circ})$ & $(0.0618, -0.0624, 0.0370)$ \\
$(1,0,0)$ & I & $(0, 0, 0,  -0.05^{\circ}); \, (\pi,\pi,\pi, 0.05^{\circ})$ & $(0.0630, 0.0636, -0.0390)$ \\
\hline
$(-1,1,1)$ & I & $(0, 0, \pi, -0.03^{\circ}); \, (\pi,\pi,0, 0.03^{\circ})$ & $(-0.0496, -0.0504, 0.0035)$ \\
$(1,-1,1)$ & N & $(0, 0, 0, -6.6^{\circ}); \, (\pi, \pi, \pi, 6.6^{\circ})$ & $(0.0078, -0.0117, -0.0501)$ \\
$(1,1,-1)$ & N & $(\pi, \pi, 0, -4.8^{\circ}); \, (0, 0, \pi, 4.8^{\circ}) $ & $(0.0032, -0.0093, 0.0496)$ \\
\hline
$(0,1,1)$ & I & $(\pi, \pi, \pi, -0.1^{\circ}); \, (0, 0, 0, 0.1^{\circ})$ & $(0.0522, 0.0530, 0.0167)$ \\
$(0,1,1)$ & I & $(\pi, \pi, 0, -0.2^{\circ}); \, (0, 0, \pi, 0.2^{\circ})$ & $(0.0522, 0.0530, 0.0167)$ \\
\hline
$(2,1,1)$ & N & $(0, 0, 0, -3.8^{\circ}); \, (\pi, \pi, \pi, 3.8^{\circ})$ & $(0.0092, 0.0127, 0.0503)$ \\
$(2,1,1)$ & N & $(0, 0, \pi, -3.6^{\circ}); \, (\pi, \pi, 0, 3.6^{\circ})$ & $(0.0210, 0.0227, 0.0538)$ \\
\hline
$(-2,1,1)$ & I & $(\pi, \pi, \pi, -0.02^{\circ}); \, (0, 0, 0, 0.02^{\circ})$ & $(-0.0498, -0.0506, 0.0055)$ \\
$(-2,1,1)$ & I & $(0, 0, \pi, -0.1^{\circ}); \, (\pi, \pi, 0, 0.1^{\circ})$ & $(-0.0523, -0.0530, 0.0168)$ \\
\hline
$(0,1,-1)$ & N & $(\pi, \pi, 0, -5.4^{\circ}); \, (0, 0, \pi, 5.4^{\circ})$ & $(0.0213, -0.0230, 0.0539)$ \\
$(0,1,-1)$ & N & $(\pi, \pi, \pi, -7^{\circ}); \, (0, 0, 0, 7^{\circ})$ & $(-0.0282, 0.0295, -0.0570)$ \\
\hline
$(1,0,1); (1,1,0)$ & N & $(\pi, 0, 0, -4.4^{\circ}); (\pi, 0, \pi, 4.4^{\circ})$ & $(0.0073, 0.0114, -0.0500)$ \\
\hline
$(1,2,1); (1,1,2)$ & N & $(0, \pi, \pi, -4.4^{\circ}); (0, \pi, 0, 4.4^{\circ})$ & $(0.0572, -0.0578, -0.0756)$ \\
\hline
$(1,-2,1); (1,1,-2)$ & N & $(\pi, \pi, 0, -4.2^{\circ}); (0, 0, \pi, 4.2^{\circ})$ & $(0.0074, -0.0114, -0.0500)$ \\
\hline
\end{tabular}
\caption{The collection of vev alignments considered in our parameter
  scan, resulting hierarchy, phases, deviation from bimaximality, and
  neutrino masses.  The mass hierarchy column indicates whether the
  vev generates a normal (N) hierarchy or inverted (I) hierarchy.  The
  last two columns indicate the required phases to obtain a valid mass
  hierarchy and the $\theta_{23}$ angle and neutrino masses
  corresponding to the central $\theta_{13}$ value from Daya
  Bay~\cite{An:2012eh}. }
\label{table:vevlist}
\end{table}

\subsection{Dirac $CP$ Phase Predictions}

\begin{figure}[t!]
\includegraphics[width=\columnwidth]{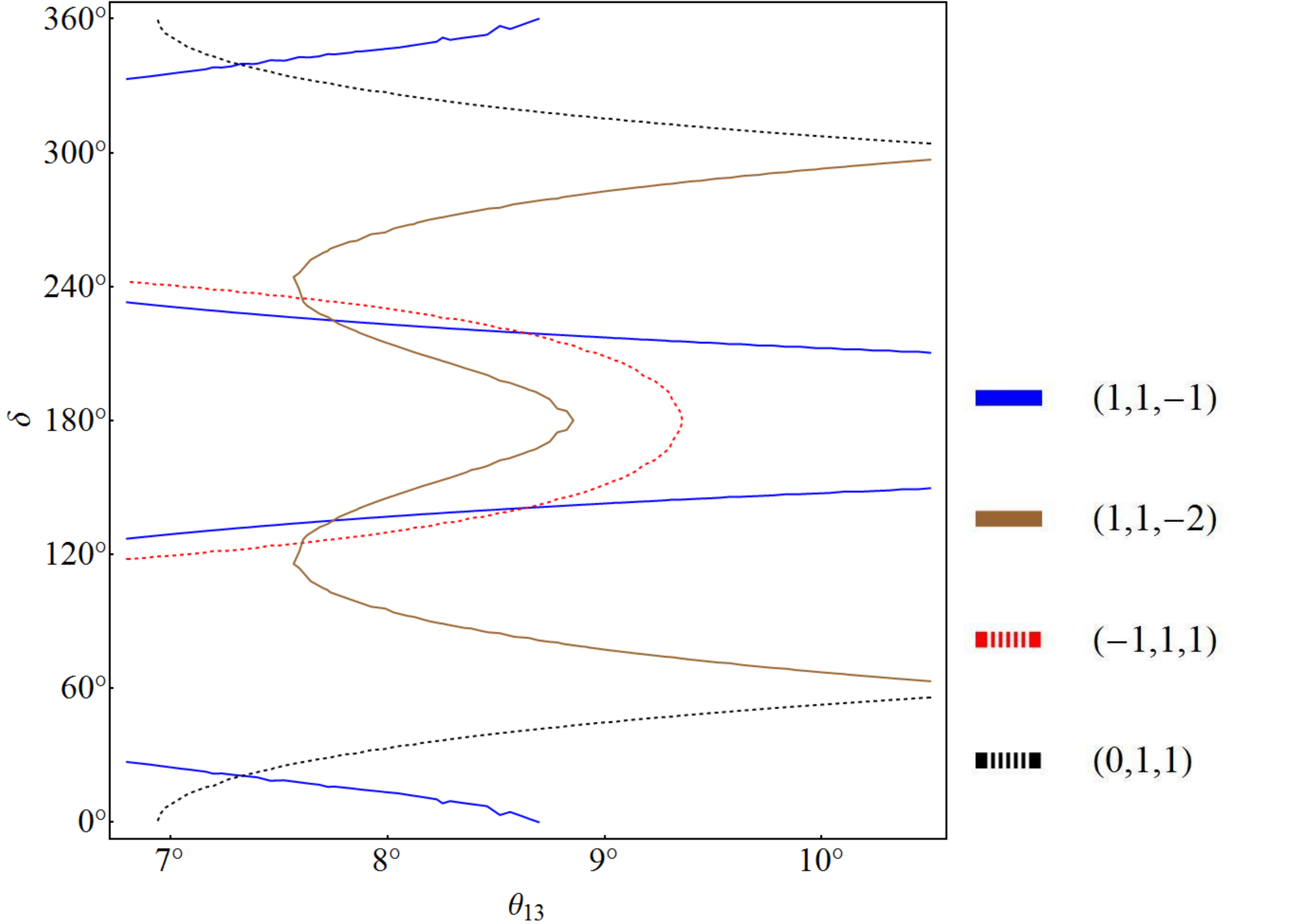} 
\caption{Contours for various $A_4$ breaking patterns at their
  respective favored $\theta_{23}$ angle based on~\tableref{vevlist}.
  The $(1, 1, -1)$ and $(1, 1, -2)$ patterns show the ratio contour
  for a normal mass hierarchy, while the $(-1, 1, 1)$ and $(0, 1, 1)$
  show the ratio for an inverted mass hierarchy.  The $(1, 1, -1)$,
  $(-1, 1, 1)$, and $(0, 1, 1)$ vevs are moderately more predictive of
  $\delta$ than the $(-1, 1, 1)$ vev.}
\label{fig:cp}
\end{figure}

We now examine the predicted Dirac $CP$ phase for various triplet vev
patterns.  Instead of fixing $\delta$ and scanning over $\theta_{23}$
and $\theta_{13}$ as before, we now set $\theta_{23}$ to the value
preferred by the central $\theta_{13}$ value of Daya
Bay~\cite{An:2012eh}, as shown in~\tableref{vevlist}, and scan this
$\theta_{23}$ slice of the $\delta$ vs. $\theta_{13}$ plane.  Contours
that satisfy the correct mass squared differences are highlighted and
shown in~\figref{cp} for a few illustrative choices of vev patterns.
Thus, for $\theta_{13}$ within the $2\sigma$ range
of~\cite{An:2012eh}, the favored range of $\delta$ can be broad, as
for the $(1, 1, -2)$ vev, or fairly narrow, as for the $(1, 1, -1)$,
$(-1, 1, 1)$ and $(0, 1, 1)$ vevs.  This figure shows that a
measurement of $\delta$ and further refinement in narrowing the
$\theta_{13}$ uncertainties can exclude or significantly favor a
particular set of $A_4$ vev patterns.  In addition, we see that shifts
in the central value of $\theta_{13}$ will serve to disfavor
particular vev patterns as well as better accommodate other vev
patterns.  Certainly, more data is needed to test these possibilities
and the $A_4$ paradigm.

\section{Conclusions}
\label{sec:conclusions}
In light of the results from the Daya Bay and RENO collaborations, we
have developed a framework for understanding the constraints on the
$A_4$ parameter space from low energy neutrino observables.  For our
parameter scan, we have assumed a Type I seesaw model with a minimal
$A_4$ flavor structure governing the charged lepton masses and the
Dirac neutrino masses, but the Majorana mass matrix has contributions
from each of the possible $\mathbf{3}$, $\mathbf{1}$, $\mathbf{1'}$,
and $\mathbf{1''}$ flavons, where the triplet vev pattern is initially
unconstrained.  We categorize the triplet vev according to $Z_3$ or
$Z_2$ subgroup preserving patterns, which enhances the model
predictivity: for these subgroup preserving patterns, four $A_4$
parameters are used to predict seven neutrino observables.  Regarding
the chosen vev patterns, we leave the question of scalar potentials or
vacuum alignment for future work.

We have analyzed the $A_4$ parameter space in two distinct and
intriguing slices.  In the first case, we fix $\theta_{12}$ to be
trimaximal and scan over the resulting $\theta_{23}$ vs. $\theta_{13}$
plane for various choices of Dirac and Majorana phases.  The results
show that vev patterns giving a normal neutrino mass hierarchy have
moderately large breaking of bimaximality for $\theta_{23}$, while
inverted hierarchies generally retain bimaximality as a prediction.
This indicates that a non-bimaximal $\theta_{23}$ measurement, such as
the preliminary result $\sin^2 (2\theta_{23}) = 0.94^{+0.04}_{-0.05}$
from MINOS~\cite{Nichol} is favorably correlated with a normal
hierarchy in our $A_4$ framework.  We also analyzed the predictions
for the Dirac $CP$ phase $\delta$ for some illustrative choices of
triplet vev pattern in the $\delta$ vs. $\theta_{13}$ plane.  This
analysis emphasizes the point that a future measurement of $\delta$,
decreased uncertainty in $\theta_{13}$, and shifts in the central
value of $\theta_{13}$ can strongly favor or exclude particular
triplet vev patterns, highlighting the fact that future experimental
results have significant power in discriminating possible $A_4$ flavon
structures.


\section*{Acknowledgements}
\label{sec:acknolwedgements}
We would like to thank Michael Ratz for useful comments. M.-C.C. would
like to thank the Technische Universit\"at M\"unchen (TUM), the
Galileo Galilei Institute for Theoretical Physics (GGI), the Center
for Theoretical Underground Physics and Related Areas (CETUP* 2012) in
South Dakota, and the UC Gump Station on Moorea for their hospitality
and for partial support during the completion of this work.
M.-C.C. and FY acknowledge the hospitality of the Aspen Center for
Physics, which is supported by the National Science Foundation Grant
No. PHY-1066293.  JH would like to thank the hospitality of the
University of Washington where part of this work was completed.  The
work of M.-C.C. and JH is supported by the National Science Foundation
under Grant No. PHY-0970173. JH is also supported by the DOE Office of
Science and the LANL LDRD program. The work of JO'B and AW is
supported by the NSF under Grant No. PHY-0970171.  Fermilab is
operated by Fermi Research Alliance, LLC under Contract
No. De-AC02-07CH11359 with the United States Department of Energy.


\begin{appendix}

\section{General VEV Solutions of Flavon Fields}
\label{sec:generalvevsolns}
In~\eqnref{MN_system}, we have six equations that relate the $A_4$
flavons comprising the Majorana mass matrix to the low energy neutrino
masses, mixing angles, and phases.  We can solve this system to
express the triplet flavon components, $(\phi_a, \phi_b, \phi_c)^T$,
and the one-dimensional flavons, $\eta$, $\chi$, and $\psi$, in terms
of the physical neutrino observables.  This solution set is
\begin{align}
\label{eqn:genphia}
\langle \phi_a \rangle & =  F \left(
\frac{1}{m_1} 
	\left[ c^2_{12} c^2_{13} - \left( s_{12} s_{23} 
	- e^{-i\delta} c_{12} c_{23} s_{13} \right) 
	\left( -c_{23} s_{12} - e^{i\delta} c_{12} s_{13} s_{23} \right) \right] 
	\right. 
\nonumber\\
&\qquad 
+ \frac{1}{m_2} e^{-i 2\xi_1} \left[ 
	c^2_{13} s^2_{12} + \left( c_{12} s_{23} 
	+ e^{-i\delta} c_{23} s_{12} s_{13} \right) \left( c_{12} c_{23} 
	- e^{i\delta} s_{12} s_{13} s_{23} \right) \right]
\nonumber \\	
&\qquad 
\left. + \frac{1}{m_3} e^{-i 2\xi_2}(s^2_{13} - c^2_{13} c_{23} s_{23}) \right) \ ,
\\
\label{eqn:genphib}
\langle \phi_b \rangle & =  F \left(
 \frac{1}{m_1} \left[ 
	\left( s_{12} s_{23} - e^{-i\delta} c_{12} c_{23} s_{13} \right)
	\left( s_{12} s_{23} - e^{ i\delta} c_{12} c_{23} s_{13} \right)
	+ c_{12} c_{13} 
	\left( c_{23} s_{12} + e^{-i\delta} c_{12} s_{13} s_{23} \right) \right]
	\right.
\nonumber\\
&\qquad 
+ \frac{1}{m_2} e^{-i 2\xi_1} \left[
	\left( c_{12} s_{23} + e^{-i\delta} c_{23} s_{12} s_{13} \right)
	\left( c_{12} s_{23} + e^{ i\delta} c_{23} s_{12} s_{13} \right)
	- c_{13} s_{12} 
	\left( c_{12} c_{23} - e^{-i\delta} s_{12} s_{13} s_{23} \right)	\right]
\nonumber \\
&\qquad \left.
+ \frac{1}{m_3} 
	e^{-i 2\xi_2}(c^2_{13} c^2_{23} - e^{-i\delta} c_{13} s_{13} s_{23}) \right) \ ,
\\
\label{eqn:genphic}
\langle \phi_c \rangle & =  F \left(
\frac{1}{m_1} \left[c_{12} c_{13} 
	\left( -s_{12} s_{23} + e^{-i\delta} c_{12} c_{23} s_{13} \right)
	+ \left( c_{23} s_{12} + e^{ i\delta} c_{12} s_{13} s_{23} \right) 
  \left( c_{23} s_{12} + e^{-i\delta} c_{12} s_{13} s_{23} \right) \right]
	\right.
\nonumber\\
&\qquad 
+ \frac{1}{m_2} e^{-i 2\xi_1} \left[ c_{13} s_{12} 
	\left( c_{12} s_{23} + e^{-i\delta} c_{23} s_{12} s_{13} \right)
	+ \left( c_{12} c_{23} - e^{ i\delta} s_{12} s_{13} s_{23} \right)
  \left( c_{12} c_{23} - e^{-i\delta} s_{12} s_{13} s_{23} \right) \right]
\nonumber \\
&\qquad \left.
+ \frac{1}{m_3} e^{-i 2\xi_2}
	(c^2_{13} s^2_{23} - e^{-i\delta} c_{13} c_{23} s_{13}) \right) \ ,
\\
\label{eqn:geneta}
\langle \eta \rangle & =  \frac{F}{3} \left(
 \frac{1}{m_1} 
 [ c_{12}^2c_{13}^2 
  + 2 ( -s_{12}^2 s_{23} c_{23} + c_{12}^2 s_{13}^2 s_{23} c_{23} 
  + e^{-i\delta} c_{12} s_{12} s_{13} c_{23}^2 
  - e^{i\delta} c_{12} s_{12} s_{13} s_{23}^2 ) ] \right.
\nonumber\\
&\qquad 
+ \frac{1}{m_2} e^{-i 2\xi_{1}}
 [ c_{13}^2 s_{12}^2 
  - 2 c_{12}^2 c_{23} s_{23} - 2 c_{12} s_{12} s_{13} 
  ( e^{-i\delta} c_{23}^2 - e^{i\delta} s_{23}^2 ) 
  + 2 s_{12}^2 s_{13}^2 s_{23} c_{23}] 
\nonumber \\
&\qquad \left.
+ \frac{1}{m_3} e^{-i 2\xi_{2}} ( s_{13}^2 + 2c_{13}^2 c_{23} s_{23} ) \right) \ ,
\\
\label{eqn:genpsi}
\langle \psi \rangle & =  \frac{F}{3} \left(
 \frac{1}{m_1} 
 [ c_{23}^2 s_{12}^2 - c_{12} c_{23} s_{13} ( 2 e^{-i\delta} c_{12} c_{13} 
 - ( e^{-i\delta} + e^{i\delta} ) s_{12} s_{23} ) 
 + s_{23} ( 2 c_{13} s_{12} c_{12} + c_{12}^2 s_{13}^2 s_{23} ) ] \right.
\nonumber\\
&\qquad 
+ \frac{1}{m_2} e^{-i 2\xi_{1}}
 [ c_{12}^2 c_{23}^2 
 - ( e^{-i\delta} + e^{i\delta} ) c_{12} c_{23} s_{12} s_{13} s_{23} 
 - s_{12} ( 2 e^{-i\delta} c_{23} s_{12} s_{13} c_{13} 
 + s_{23} ( 2 c_{12} c_{13} - s_{12} s_{13}^2 s_{23} ) ) ] 
\nonumber \\
&\qquad \left.
+ \frac{1}{m_3} e^{-i 2\xi_{2}} 
	[ c_{13} ( 2 e^{-i\delta} c_{23} s_{13} + e^{i\delta} c_{13} s_{23} ) ]
	\right) \ ,
\\
\label{eqn:genchi}
\langle \chi \rangle & =  \frac{F}{3} \left(
 \frac{1}{m_1}
 [ -2 c_{13} c_{23} s_{12} c_{12} 
 - ( e^{-i\delta} + e^{i\delta} ) c_{12} c_{23} s_{12} s_{13} s_{23} 
 + s_{12}^2 s_{23}^2  - c_{12}^2 
 ( -c_{23}^2 s_{13}^2 + 2 e^{-i\delta} s_{13} c_{13} s_{23} ) ] \right.
\nonumber \\
&\qquad 
+ \frac{1}{m_2} e^{-i 2\xi_{1}}
 [ c_{12}^2 s_{23}^2 
 + s_{12}^2 s_{13} ( c_{23}^2 s_{13} - 2 e^{-i\delta} c_{13} s_{23} )  
 + c_{12} c_{23} s_{12} 
 ( 2 c_{13} + ( e^{-i\delta} + e^{i\delta} ) s_{13} s_{23} ) ] 
\nonumber \\
&\qquad \left.
+ \frac{1}{m_3} e^{-i 2\xi_{2}} 
	[ c_{13} ( c_{13} c_{23}^2 + 2 e^{-i\delta} s_{13} s_{23} ) ] \right) \ ,
\end{align}
where $F = \dfrac{ v_H^2 \lambda_N^2 }{\Lambda_{RR}}$.  We
invert~\eqnref{genphia},~\eqnref{genphib}, and~\eqnref{genphic} to
solve for the neutrino masses in terms of the triplet vev components
$(\phi_a, \phi_b, \phi_c)^T$.  Then, we constrain the one-dimensional
flavons and the neutrino masses by assuming a particular triplet vev
pattern and scanning over the mixing angles and phases, applying the
constraint on mass squared differences from~\cite{Schwetz:2011qt} and
the cosmological bound on the sum of absolute neutrino masses
from~\cite{Nakamura:2010zzi}.

\end{appendix}

\bibliographystyle{apsrev}
\bibliography{./theta13}



\end{document}